\documentclass[a4paper,11pt]{article}
\usepackage{amsfonts}
\usepackage{latexsym}
\usepackage{amsmath}
\usepackage{amssymb}
\usepackage{amssymb}
\usepackage{slashed}
\usepackage{upgreek}
\usepackage{mathrsfs}
\usepackage{bbm}

\usepackage{tikz}

\usepackage{tikz-3dplot}

\usepackage{amsthm}

\theoremstyle{remark}

\newtheorem*{rmk*}{Remark}

\theoremstyle{remark}

\usepackage[utf8]{inputenc}
\usepackage[english]{babel}

\theoremstyle{definition}

\allowdisplaybreaks

\usepackage{relsize}
\usepackage{graphicx}
\usepackage{comment}

\renewcommand{\thefootnote}{\fnsymbol{footnote}}

\def\appendix#1{\addtocounter{section}{1}\setcounter{equation}{0}
\renewcommand{\thesection}{\Alph{section}}
\section*{Appendix \thesection\protect\indent \parbox[t]{11.15cm}{#1}}
\addcontentsline{toc}{section}{Appendix \thesection\ \ \ #1}}

\font\mybb=msbm10 at 11pt

\def\bb#1{\hbox{\mybb#1}}

\def\bR {\bb{R}}

\def\rsq {]\kern -2.0pt]}
\def\lsq {[\kern -2.0pt[}

\def\be{\begin{equation}}
\def\ee{\end{equation}}

\def\rsq {]\kern -2.0pt]}
\def\lsq {[\kern -2.0pt[}

\newcommand{\bea}{\begin{eqnarray}}
\newcommand{\eea}{\end{eqnarray}}

\usepackage{microtype}
\usepackage[colorlinks, citecolor=blue, linkcolor=blue, urlcolor=Maroon, filecolor=Maroon, linktocpage=true]{hyperref} 

\usepackage{chngcntr}
\counterwithout{equation}{section}

\begin{document}

\begin{center}
\vspace*{-1.0cm}
\begin{flushright}
\end{flushright}




\vspace{2.0cm} {\Large \bf Sigma model renormalisation group flows,  singularities and some remarks on  cosmology} \\[.2cm]



\vskip 2cm
 Georgios  Papadopoulos
\\
\vskip .6cm


\begin{small}
\textit{Department of Mathematics
\\
King's College London
\\
Strand
\\
 London WC2R 2LS, UK}\\
\texttt{george.papadopoulos@kcl.ac.uk}
\end{small}
\\*[.6cm]

\end{center}

\vskip 2.5 cm

\begin{abstract}
\noindent

We  investigate the properties of the renormalisation group (RG) flow of two-dimensional sigma models  with a generic metric coupling by utilising  known results for the Ricci flow. We point out that on many occasions the RG  flow develops singularities, due to strong coupling behaviour, before it reaches  a UV or an IR fixed point.  We illustrate our analysis with several examples. We give particular emphasis to type I singularities, where the length of the curvature of the sigma model target space grows at most as $|t-T|^{-1}$ as the flow parameter $t$ approaches the singularity at $T$.  For these, the geometry near the singularity is described in terms of a  shrinking  Ricci soliton that exhibits a cosmological constant even though the original RG flow does not.

Assuming that the  spacetime satisfies an RG flow equation, we use  the Ricci solitons  to introduce a cosmological constant in a string theory setting. This can allow for   different cosmological constants at different regions of spacetime.  In particular, we point out how the de-Sitter space is a solution  of the theory. We also raise the question on whether the techniques used to prove the geometrisation conjecture can be applied to prove the homogeneity and isotropy of the universe at large scales.

\end{abstract}



\newpage

\renewcommand{\thefootnote}{\arabic{footnote}}



\numberwithin{equation}{section}

\section{Introduction}\label{intro}

 One objective of this letter is to give a description of renormalisation group (RG) flow of a two-dimensional sigma model, introduced by Friedan \cite{friedan}, starting from a generic metric $g$ on the sigma model manifold $M^n$. This is based on the analysis of Ricci flows as developed by Hamilton \cite{Hamilton, Hamilton0} and Perelman \cite{Perelman, Perelman1, Perelman2}. It will be pointed out that on many occasions, the RG flow exhibits singularities before it reaches an ultraviolet (UV) or a infrared (IR) fixed point, where the sigma model perturbation theory breaks down because of strong coupling effects. Several examples of RG flow will be given including  flows that develop type I singularities and the role of Ricci solitons will be elucidated.

 The second objective is to point out an application to cosmology. This includes  the emergence of a cosmological constant in a string theory setting and potentially the construction of a proof for the isotropy and homogeneity of our universe at large scales. The second objective  requires a number of hypotheses  and it may not be realistic but, nevertheless, it can be proved useful as a set up, a toy model,  that can be used to explore  such questions.

The RG equation of two-dimensional sigma models has extensively been explored  for several reasons. One reason is that some sigma models, like the $O(N)$ sigma models, whose target space is the $S^{N-1}$ sphere, for $N>2$ are asymptotically free at the UV, and exhibit a strong coupling behaviour at the IR, for a review see \cite{novikov}. Thus they have been investigated as a paradigm for QCD. In this context, this and other similar models are assumed  to have a high degree of symmetry. The RG equation determines the running of a finite number of coupling constants, usually just the radius of the (sigma model) target space, that are allowed by the symmetries of the sigma model.

Another reason for studying two-dimensional sigma models is that they have found applications in the context of string theory -- in particular, the construction of low energy effective theories for strings \cite{CFMP, Ts, Ts2, CT}.  In string theory, the focus is on the fixed points of the sigma model RG equation described by two-dimensional conformal field theories, instead on the RG  flows themselves. However,  the novelty in this context is that sigma model perturbation theory was developed for generic metrics $g_{\mu\nu}$ on the sigma model target space  $M^n$ \cite{friedan, Honerkamp, Howe} instead of special ones that enjoy a high degree of symmetry.  In particular, one of the key results that emerged from the applications of sigma models to strings is  that for a bosonic two-dimensional sigma model\footnote{The fields $\phi$ of a two-dimensional sigma model are maps from the two-dimensional Minkowski space with metric $\eta_{ab}$, $a,b=0,1$, into the target space $M^n$ with metric $g_{\mu\nu}$.} with  Lagrangian
\be
L=\frac{1}{2}\, g(\phi)_{\mu\nu}\,\eta^{ab} \partial_a \phi^\mu\, \partial_b \phi^\nu~,
\ee
the RG equation  at one-loop \cite{friedan} can be written as
\be
\frac {d g(t)_{\mu\nu} }{dt}= -  R(g(t))_{\mu\nu}~,
\label{fl1}
\ee
where $R_{\mu\nu}$ is the Ricci tensor of $g_{\mu\nu}$ and $ t=-\log \mu$ with $\mu$ the two-dimensional energy scale.  Of course, the RG flow equation for bosonic sigma models is expected to be   corrected to all loops. However, here we shall focus on investigating the one loop contribution as this is the leading order correction and, moreover, there is a limited analytic information about the solutions of this  equation if higher order curvature corrections are included in the right hand side.

RG flows in two-dimensional field theories have been extensively investigated. Zamolodchikov \cite{Zam} proved the monotonicity of RG flows for two-dimensional field theories by demonstrating that there is a function that decreases during the flow and at the fixed points is identified with the central charge of the associated conformal theory. Polchinski \cite{Polchinski} showed that under certain assumptions all scale invariant two-dimensional field theories are actually conformally invariant. These results have been adapted to sigma models with $B$-field coupling  in \cite{OSW2, OSW, Tseytlin2, Huhu}. Remarkably to include the $B$-field coupling,  it involves an extension  of the work of Perelman. The  conditions required for two-dimensional sigma models to be   scale and conformal invariant  have been investigated  in \cite{HullTownsend}. In \cite{GPEW} has been demonstrated that all scale invariant sigma models with compact, or at least geodesically complete, target space  are conformally invariant -- there is an analogous result in generalised Ricci flows  proven in \cite{GFS}.   This deep relation between Ricci flows and their generalisations with  sigma model  RG flows underpins our constructions.

The question that will be explored  is the following: what is the behaviour of RG flows as either $\mu\rightarrow 0$ (equivalently $t\rightarrow +\infty)$) or $\mu\rightarrow +\infty$ (equivalently $t\rightarrow -\infty$)?  The emphasis  will be  on the former, i.e. what does it happen to the metric $g(t)$ as the energy scale decreases  starting from a generic metric $g_0$ on the sigma model manifold $M^n$?  In particular, suppose that one fixes the topology of $M^n$ and starts at a scale $\mu_0$ with a generic metric $g_0$, how does one expect that $g_0$ will evolve with (\ref{fl1})  as the energy scale goes into infrared? The analogous question with the scale going into UV can be analysed in a similar way.  This also applies to manifolds $M^n$ that potentially admit a high degree of symmetry, like the spheres, but the starting point can be a metric $g_0$ that may not admit any isometries -- the behaviour of the RG flow depends on the initial metric $g_0$. {\it We find that generically the sigma model RG flow (\ref{fl1}) reaches strong coupling singularities, where the curvature of the sigma model manifold diverges, instead of converging smoothly to a UV or IR fixed point.}

The description of the Ricci flows, and consequently that of the sigma model RG flows, will focus on the singularities that can develop at some value of the flow parameter $t$. After some examples,  the type I singularities will be examined in more detail. For this the length of the Riemann curvature grows at most as $|t-T|^{-1}$ as $t$ approaches the value $t=T$. It has been pointed out  that the geometry of $M^n$ near such a singularity is described by a Ricci soliton \cite{Perelman, Naber, Enders}. It will be emphasised that Ricci solitons can be seen in two different ways. One way is  as solutions of Einstein like equations {\it with potentially a cosmological constant term} -- see (\ref{steadsol}) below -- or as solutions of the flow
equation (\ref{fl1}) that {\it does not exhibit a cosmological constant}.  It is this dual nature of the Ricci solitons that will be utilised for the hypothesis that this set up can be used as a toy model for cosmology in a string-like theory setting.

After the Big Bang,   inflation \cite{guth, linde}  provides an  explanation for the homogeneity and isotropy of the universe at  very large scales -- this is also known as the horizon problem. Within the inflationary scenario,  it is proposed that all the observable universe today is the result of the exponential expansion of a small region in the early universe that was in causal contact -- this explains why the temperature of the microwave radiation is almost uniform in every direction.  Forward to present times,
it is  widely accepted that the universe  has a small positive, but not vanishing, cosmological constant, $\Lambda\sim 10^{-123} l^{-2}_{\mathrm{Pl}}$,  that drives its expansion. The difficulty of incorporating such a small cosmological constant in field theory and a cosmological constant in string theory, is  well documented, for string theory see the proposals of e.g. \cite{KKLT} and \cite{Obied}. The applications of string theory to cosmology has been initiated in \cite{nappi} and for a more recent works see e.g. \cite{jcdk} and references therein.  Other approaches to cosmology based on string theory include the holographic cosmology \cite{witten3, strominger1, strominger2, maldacena}, see also \cite{bvv, mcfs, HNKS} for the mergence of time in this context as the renormalisation group flow parameter of the boundary theory.  An early work on the use of RG flows in string cosmology is presented  in \cite{bakas}.

Here, we point out that spacetimes with non-vanishing positive cosmological constant can emerge as solutions of the RG flow equation (\ref{fl1}). Therefore, under  the hypothesis that the universe satisfies an RG flow equation as in (\ref{fl1}),  spacetimes with a positive cosmological constant can be incorporated in a string theory-like  setting.  We shall  achieve this by utilising the properties of the Ricci solitons  that are solutions to the sigma model RG flow.  In particular, we shall show that   de-Sitter space is such a solution of (\ref{fl1}).  This is the case even though (\ref{fl1}) does not exhibit a cosmological constant and the de-Sitter space does. Moreover, as Ricci solitons can model the geometry of the spacetime near certain RG flow singularities\footnote{RG flow singularities should not be confused with spacetime singularities. Although both lead to a singular spacetime  their origin and the mechanism of formation are very different.} and these singularities can occur at different regions of the spacetime,  we shall hypothesise that potentially different cosmological constants can occur at different epochs.


It has already been mentioned that the inflationary scenario  \cite{guth, linde} provides an explanation for the homogeneity and isotropy of the universe at very large scales.  As a concluding remark, it is worth pointing out that the scenario of staring with a universe that exhibits a generic geometry after the Big Bang and later evolves to a homogeneous and isotropic one is  reminiscent to the question answered  in the Perelman's proof \cite{Perelman2} of Thurston's  geometrisation  conjecture \cite{Thurston} in the context of geometric flows. There is no evidence that the two mechanisms are related or the later approach can be applied to prove the homogeneity and isotropy of universe. But if it does, it will provide an alternative geometric approach into the problem.

This letter is organised as follows. In section two, we begin with a summary of the solutions to the Ricci flow based on examples. We describe the role of Ricci solitons and their relation to Ricci flow singularities. The type I singularities are  introduced and some of their properties are explained   in detail. In the second part of the section, we apply the Ricci flow results to the sigma model RG flows. {\it It is emphasised that on many occasions, the RG flow reaches a strong coupling singularity instead of converging smoothly to a fixed point in either the UV or the IR.}  In section three, we apply some of our conclusions to cosmology. In particular, we point out that spacetimes, like de-Sitter space, are solutions to the sigma model RG flow. Moreover, type I singularities can provide a mechanism to introduce  different cosmological constants at different regions of spacetime. We conclude with a discussion regarding the homogeneity and isotropy of the universe and raise the question on whether this can be proven with similar techniques as those used in the proof of the geometrisation conjecture.

\section{Renormalisation group flows, Ricci solitons and singularities}

To investigate the behaviour of RG flows, we shall first summarise some of the results known about the behaviour Ricci flows. There is a vast and constantly evolving literature that explores the properties of Ricci flows   and in particular their singularities. Here, the focus will be on developing an intuition in dealing with them and the description of some simple examples that later will be useful to explain the type I singularities.

\subsection{Ricci flows and singularities}

\subsubsection{Ricci solitons and singularities}

There are two aspects of Ricci flows that they will be the subjects of our exploration. One is a qualitative description of the singularities that can occur. It is known  that on many occasions that Ricci flows develop a singularity at some finite value of the flow parameter $t$. As it has already been mentioned, the focus will be on some  simple examples. The only general case of singularities we shall describe are the so called type I singularities. The second aspect is the Ricci solitons.  These satisfy the condition
\be
R_{\mu\nu}=\Lambda g_{\mu\nu}+\nabla_\mu X_\nu+\nabla_\nu X_\mu~,
\label{steadsol}
\ee
where $\Lambda$ is a constant\footnote{We shall often refer to $\Lambda$ as the cosmological constant. However, in $n>2$ dimensions the cosmological constant, $\Lambda_c$, as it  appears in the Einstein equations, $G+\Lambda_c g=\kappa T$,  is related to $\Lambda$ as $\Lambda=\frac{2 \Lambda_c}{n-2}$, where $G$ is the Einstein tensor.} and $X=X^\mu\partial_\mu$ is a vector field on $M^n$ with $X_\mu=g_{\mu\nu} X^\nu$. The solitons with  $\Lambda<0$, $\Lambda=0$ and $\Lambda>0$ are {\it expanding}, {\it steady} and {\it shrinking}, respectively.  Moreover,  if $X=-d\Phi$, then the Ricci solitons are called {\it gradient} solitons. As we shall mention below, type I singularities in the Ricci flow are related to  Ricci solitons -- the latter provide a model for the geometry of the manifold $M^n$ near the type I singularities.

Ricci solitons can be seen in two different ways. One definition is that mentioned above  as solutions to (\ref{steadsol}).  Alternatively,
 they are special solutions to the flow equation\footnote{In Ricci flows, the flow equation is written with an additional factor of two, schematically, as $\frac{d}{dt} g=-2 R$. Keeping this notation introduces  additional  factors of two in equations and so for simplicity we shall use (\ref{fl1}) instead.} (\ref{fl1}) of the type $g(t)= \Omega^2(t) f^*_t g_0$, where $\Omega^2$ is a scaling of the metric that depends only on $t$ and $f_t$ is a family of diffeomorphisms of $M^n$.  Assuming that the family of diffeomorphisms $f_t$ is generated by the vector field $Y(t)$, then $\Lambda=-d\Omega^2/dt\vert_{t=0}$ and $X=-Y(0)$. Conversely given a solution to (\ref{steadsol}), one can construct a solution of the Ricci flow equation (\ref{fl1}),
 \be
 g(t)= \Omega^2(t) f^*_t g_0~,
 \label{soflow}
 \ee
  with
\be
\Omega^2(t)=1-\Lambda t~,~~~Y_t(x)=-\Omega^{-2} X(x)~.
\ee
See e.g. \cite{BCDK}, page 23,    for the proof.
There are two curiosities with  flows associated to Ricci solitons. One is the presence of a cosmological constant term in (\ref{steadsol}), even though the flow (\ref{fl1}) is just a Ricci flow without a cosmological constant.  The other curiosity is that for gradient solitons with $\Lambda=0$, the resulting equation
\be
R_{\mu\nu}+2 \nabla_\mu \partial_\nu \Phi=0~,
\label{grad}
\ee
is that required for a sigma models to be conformally invariant. This means that such an equation can either describe a RG flow generated by $X=-d\Phi$ or the condition for conformal invariance for a sigma model with only a metric coupling.

{\it Example 1:}~~The simplest example\footnote{This and other examples that we present below are not new, see e.g. \cite{Hamilton0, BCDK}. However here, in addition to their usual expression in the literature as solution to (\ref{steadsol}),   we  include the explicit expression of the Ricci solitons as solutions to the Ricci flow equation (\ref{fl1}).}   of a gradient shrinking Ricci soliton with $\Lambda>0$ is that for which $g$ is the flat metric on $\bR^n$ and  $\Phi=\frac{\Lambda}{4} x^2$, see e.g. \cite{BCDK}. Then $\Omega^2=(1-\Lambda t)$ and
$Y_t(x)= -\frac{\Lambda}{ 2 (1-\Lambda t)} x^\mu\partial_\mu$.  The flow of the vector field $Y_t$ is given by the curves $x^\mu(t)=\frac{1}{\sqrt{1-\Lambda t}} x_0^\mu$ for $t\in (-\infty, \frac{1}{\Lambda})$, where the boundary condition has been imposed at $t=0$.  Thus, using the flow $f_t$ of $Y_t$ and (\ref{soflow}), we find that
\be
g(t)=(1-\Lambda t) f_t^* g_0= g_0~,
\ee
as expected.

{\it Example 2:}~~The standard paradigm is that of the flow of Einstein metrics. For such metrics, the Ricci tensor is expressed in terms of the metric  $R_{\mu\nu}= \Lambda g_{\mu\nu}$, where $\Lambda$ is a constant. Next consider solutions of the Ricci flow equation (\ref{fl1}) of the form $g(t)=\Omega^2(t) g$. This gives
 \be
 g(t)_{\mu\nu}= -\big (\Lambda t- 1 \big ) g_{\mu\nu}~,~~~
 \ee
 where $t_0=0$ and $g(0)=g$ -- if $t_0\not=0$, then $g(t)= -(\Lambda (t-t_0)-1) g(t_0)$.
 Clearly for $\Lambda>0$, the solution is smooth for $t<T=\Lambda^{-1}$ and a singularity develops at $t=T$ in the ``future''. Clearly, the solution is valid for all $-\infty<t<T$ -- in Ricci flow terms,  this is called an {\it ancient} solution.  For $\Lambda<0$, the solution is smooth for all $t>-T$, where $T=-\Lambda^{-1}$ but it develops a singularity at $t=-T$.  Thus this solution is well defined in the ``future'' but has a singularity in the ``past''  -- in Ricci flow terms, this is an {\it immortal} solution.  As we shall see below, there are solutions that are smooth for all $t$. In Ricci flow terms, these  are called {\it eternal} solutions. These are the only ones with this property as all the others exhibit a singularity either in the past or the future.

To  provide a qualitative description of the  singularities that develop in the solutions of the flow equations (\ref{fl1}),
 the behaviour of Einstein metrics under geometric flow is indicative of what it is happening for generic metrics on manifolds. A broad stroke conclusion is that under Ricci flows as $t\rightarrow +\infty$  the manifold along directions of positive curvature contracts while along negative curvature directions expands.

{\it Example 3:}~~ The typical example for this explained in \cite{Hamilton0} is that of the dumbbell  -- a rod (neck) that smoothly joins two spheres at the two ends. Topologically a $n$-dimensional dumbbell is a $S^n$ sphere equipped with a different metric from the usual round one.  For a two-dimensional dumbbell, the neck is a cylinder $S^1\times (-1,1)$. As the curvature of $S^1$ vanishes, the neck is slightly negatively curved.  As a result it is expected that under  Ricci flow with $t\rightarrow +\infty$ the diameter of the neck to increase and ``inflate'' the dumbbell to a round $S^2$.  However, for any n-dimensional dumbbell, with $n>2$,  the positive curvature of the $S^{n-1}$ sphere section at the neck of the dumbbell will ``win'', provided that the spheres at the two ends are less curved, and the neck will shrink under the flow and develop a pinch at some finite $t=T$.
  This is a singularity, a {\it neck pitch} singularity -- the Ricci   flow cannot run pass through  it.  Typically, to continue the geometric flow pass the singularity
  a non-trivial procedure is required, like for example surgery, to repair the singularity and then start again the flow.

{\it Example 4:}~~Finally, we shall give an example of an eternal Ricci soliton. This is the Witten's  black hole solution \cite{Witten2} given by
 \be
 g=\frac{dx^2+dy^2}{1+x^2+y^2} =\frac{1}{1+r^2} \big(dr^2+ r^2 d\theta^2\big)
 \ee
 where $(x, y)\in \bR^2$, $r\in (0, +\infty)$ and $\theta\in (0, 2\pi)$.  Changing coordinates as $r=\sinh u$, $u\geq 0$, the metric can be rewritten as
 \be
 g= du^2+\tanh^2u\, d\theta^2~.
 \ee
   The metric is smooth on $\bR\times S^1$ -- observe that as $u\rightarrow 0$, it becomes the flat metric on the 2-plane. This metric satisfies (\ref{grad}) with
 \be
e^{-2 \Phi}= \cosh u~.
 \ee
 Thus, this is an example of a gradient steady Ricci soliton. As it has already been explained, this soliton gives rise to a solution of the Ricci flow equation  (\ref{fl1}). In this case, the Ricci flow  is generated by the vector field
 \be
 X=-\tanh u\, \partial_u~.
 \ee
 In particular, the flow $f_t$ of this vector field is
 \be
 \sinh u(t)=e^{-t} \sinh u_0~,~~~\theta=\theta_0~,
 \ee
 where the boundary condition has been imposed at $t=0$. The solution to the Ricci flow equation (\ref{fl1}) reads
 \be
 g(t)=f_t^* g_0=du^2+\tanh^2u(t)\, d\theta^2=\frac{e^{-2t} \cosh^2u_0}{1+e^{-2t} \sinh^2u_0} \Big(du_0^2+\tanh^2u_0\, d\theta_0^2\Big)~.
 \ee
 The Ricci flow $g(t)$ is smooth for $t\in (-\infty, +\infty)$.  As $t\rightarrow -\infty$, it approaches the flat metric on the cylinder.  But at $t=+\infty$, it appears to be singular. However, the singularity can be removed by setting $r= e^{-t} \sinh u_0$ in which case it describes the flat metric on the 2-plane.

\subsubsection{Type I singularities}

Type I singularities are the most well investigated and common class of Ricci flow singularities. Suppose that there is a solution to the Ricci flow equation (\ref{fl1}), which is smooth in the interval $[0, T)$,  with a singularity developing at $t=T$. At the singularity, the length of the curvature of the manifold diverges. The singularity is of type I  provided that
\be
 \sup_{M^n\times [0, T)} \big((T-t) |R(x,t)|\big)\leq C<\infty~,
 \ee
for some constant $C$, where $|R(x,t)|=\big(R_{\mu\nu\rho\sigma}(g(x,t)) R^{\mu\nu\rho\sigma}(g(x,t))\big)^{1/2}$ is the pointwise length of the Riemann tensor and the inner product is taken with respect to the metric $g(t)$.  This means that the length of the Riemann tensor does not grow faster than $(T-t)^{-1}$ as the singularity at $t=T$ is approached\footnote{It can be shown \cite{Hamilton0}  that the length of the curvature tensor grows at least as fast as $(T-t)^{-1}$ as the singularity at $t=T$ is approached.}. Clearly, type I singularities include the shrinking singularities of Einstein manifolds with positive cosmological constant. They also include some neck pinching singularities but not all. In particular, they include the dumbbell neck pinching singularities provided that the radii of the two spheres at the ends of the dumbbell are larger than that of the neck. So the neck pinches first before the spheres at the two ends shrink to a point. However, there is a situation of dumbbell neck pitching singularity for which the radius of one of the spheres at one of the ends of the dumbbell is comparable to that of the neck with the sphere at the other end to have a bigger radius than the other two. In such a scenario, the sphere at the neck and the small sphere at the end of the dumbbell shrink at about the same rate resulting to a cusp like singularity at the remaining  sphere, see e.g. \cite{BCDK} page 62 and references therein. Such a singularity will not be of type I as the curvature will blow up faster than $(T-t)^{-1}$.

One of the key properties of the type I singularities is that it is possible to ``zoom in'' and determine the behaviour of the geometry of $M^n$ near the singularity. This is achieved by choosing a sequence of points $(x_i, t_i)$ in $M\times [0, T)$ such that $t_i$ converges to $T$ from below and
\be
|R(x_i, t_i)|=\sup_{M\times [0, t_i)} |R(x, t)|
\ee
i.e. consider the points $(x_i, t_i)$ in $M\times [0, t_i)$ that the length of the curvature is maximised. Setting $Q_i=|R(x_i, t_i)|$, consider the metric
\be
g_{i \mu\nu}(s, x)= Q_i\, g_{\mu\nu}(t_i+\frac{s}{Q_i}, x)~,
\label{remetric}
\ee
for $s\in [-Q_i t_i, 0]$. Clearly, $Q_i\rightarrow \infty$ as $t_i\rightarrow T$ and the curvature of $g_i$ at $s=0$ and $x=x_i$ has been normalised to 1. In terms of the original parameter $t$, $g_i$ is defined in the interval $[0, t_i]$. Notice that with respect to the metric $g_i$, we zoom in by a factor of $\sqrt{Q_i}$ and speed up the flow by a factor of $Q_i$. To see this, a length $L$ as measured with respect to $g$ becomes $\sqrt{Q_i} L$ as measured with respect to $g_i$. Therefore, we measure bigger lengths with respect to $g_i$ as compared to those measured with respect to $g$, so we zoom in. Also as $t=t_i+\frac{s}{Q_i}$, the $s$ parameter `travels'' $Q_i$ as fast as $t$.

In the limit that $i\rightarrow \infty$, a (subsequence) $(M^n, g_i; x_i, t_i)$ converges in the Cheeger-Gromov \cite{CG} sense\footnote{Under certain assumptions that include a bound on the curvatures of $g_i$, a (sub)sequence of manifolds and metrics converges to a  smooth  metric at the limit. This limit is designed to described the local geometry around $x_\infty$. From now on, the reference point $x_\infty$ will be neglected.} as $(M^n, g_i; x_i, t_i)\rightarrow (\tilde M^n, g_\infty;  x_\infty)$ to a smooth ancient solution $\tilde M^n \times (-\infty, 0]$. Moreover, it has been demonstrated by Perelman, Naber and Enders et.al. \cite{Perelman, Naber, Enders} that if $M^n$ is complete,  $g_\infty$ is an shrinking Ricci soliton\footnote{It is remarkable that a cosmological constant arises by probing the geometry of a type I singularity.}, i.e. a solution to (\ref{steadsol}) with $\Lambda>0$.
It should be stressed though that $(\tilde M^n, g_\infty)$ is not a resolution of the Ricci flow type I singularity of the metric $g(t)$ -- though the procedure described is characterised in the literature as ``blow up''. The original flow remains singular and $(\tilde M^n, g_\infty)$  is singular for $s>0$. However, $(\tilde M^n, g_\infty)$ gives a description of the geometry of $(M^n, g(t))$ as $t$ approaches the singularity at $t=T$.

{\it Example 5:}~~Let us illustrate the construction with an example.  Suppose that $M^n=S^n$, $n\not=1$,  equipped with the round metric $g$ with $R_{\mu\nu\rho\sigma}=\frac{\Lambda}{n-1} (g_{\mu\rho} g_{\nu\sigma}- g_{\nu\rho} g_{\mu\sigma})$. Of course, this example is straightforward but nevertheless illustrates the construction. We have demonstrated that the solution of the Ricci flow (\ref{fl1}) for this metric is
$g(t)=-(\Lambda t-1) g$.  It is clear that
\be
|R(x, t)|=-\sqrt{\frac{2n}{n-1}} \frac{\Lambda}{\Lambda t-1}
\ee
and thus
\be
Q_i=-\sqrt{\frac{2n}{n-1}} \frac{\Lambda}{\Lambda t_i-1 }
\ee
with $t_i\rightarrow 1/\Lambda$.  Therefore,
\be
g_i(s, x)= Q_i g(t_i+\frac{s}{Q_i})=-Q_i  \big (\Lambda (t_i+\frac{s}{Q_i})-1\big ) g(x)=-\big(-\sqrt{\frac{2n}{n-1}}\Lambda +s\big) g(x)= g_\infty(s, x)
\ee
Clearly, $g_\infty$ is smooth on $S^n\times (-\infty, 0]$ and so it is ancient solution as expected but develops a singularity at $s=\sqrt{\frac{2n}{n-1}}\Lambda$ in the future. The cosmological constant of $g_\infty$ is normalised to 1.

The analysis of future type I singularities can be easily adapted to those that occur in the past, i.e. to singularities that arise at the regions of $M^n$ that have locally negative curvature and time $t$ is running backwards. For example, if the solution to Ricci flow is smooth at $(T, 0]$ and singular at $t=T$, then the singularity is of type I provided that
\be
 \sup_{M^n\times (T, 0]} \big((t-T) |R(x,t)|\big)\leq C<\infty~,
 \ee
 for some constant $C$. Zooming in at the neighbourhood of the singularity using similar steps as above, the solution at the limit $(\tilde M^n, g_\infty)$ will be an immortal solution to the Ricci flow defined smoothly on $[0, +\infty)\times \tilde M^n$ and it will be an expanding Ricci soliton.

\subsection{Application to sigma model RG flows}

The results we have described for the behaviour of the solutions to the  Ricci flow can be easily adapted to describe the behaviour of RG flow in two-dimensional sigma models.  As it has already been mentioned the Ricci flow parameter $t$ is related to the energy scale $\mu$ of the two-dimensional sigma models with $t=-\log\mu$. From the string perspective, the collective\footnote{Sigma models have infinite many couplings given by the components of the spacetime metric at every point of the spacetime or equivalently sigma model manifold $M^n$. The ratio $\ell_s/R_M$ provides an overall description of the couplings of the sigma model as the two-dimensional energy scale changes.} sigma model coupling is the ratio $\ell_s/R_M$ of string length $\ell_s$ with the radius $R_M$ of the sigma model manifold $M^n$.  Therefore, at strong sigma model coupling $\ell_s>>R_M$. Independently on whether this happens at the UV or IR, the (length of the)  curvature of the spacetime blows up and so the description of the string effective theory\footnote{Here, we assume an off-shell formulation of string theory away from the usual description that the sigma model is taken to be a conformal field theory, for a review see \cite{AW}.}  breaks down.

 With these identifications, sigma models on Einstein manifolds with positive cosmological constant are expected to have a UV fixed point at infinity described by a flat metric, as $|R(x, t)|\sim |(\Lambda t-1)|^{-1}$ with $t\rightarrow +\infty$.   But they exhibit IR strong coupling behaviour at  $t=T=\Lambda^{-1}$. At the infrared singularity, the curvature of the sigma model manifold blows up and  the sigma model perturbation theory breaks down. Thus in perturbation theory,  there is not a smooth interpolation between an UV and IR fixed points -- though this does not preclude such a smooth interpolation in a  non-perturbative formulation of the theory.

Similarly, sigma models on Einstein manifolds with negative cosmological constant have a good IR behaviour with a fixed point described by a  flat metric at $t=+\infty$. However, they exhibit strong coupling UV behaviour at $t=T=\Lambda^{-1}$. Again near the UV singularity, sigma model perturbation theory breaks down and there is no smooth interpolation between UV and IR fixed points.

Another more general class of RG flows are given by the expanding or shrinking Ricci solitons. Given a solution of the Ricci soliton equation (\ref{steadsol}), one can construct a solution of (\ref{fl1}) as in (\ref{soflow}).  These again will develop a strong coupling region either at the IR (shrinking $\Lambda>0$) or at the UV (expanding $\Lambda<0$).

Of course for eternal flows, the RG flow for two dimensional sigma models interpolates smoothly between UV and IR fixed points. Still, there may exist singularities as the RG flow reaches  the  UV or IR fixed points at $t=\pm\infty$.   In this case, sigma model perturbation theory can be used to analyse the theory for every  $t\in (-\infty, +\infty)$ as the effective coupling remains bounded as well as the curvature of the spacetime and its derivatives.

Next suppose that the sigma model RG flow develops a type I singularity in the IR -- the discussion for UV singularities is analogous. The limiting process  of zooming in at the singular region  from the string  perspective requires a modification.  As the quantities $Q_i$ used to rescale the metric in (\ref{remetric}) are dimension full, it is more convenient to deal with dimensionless quantities. For this, we replace $Q_i$ with $\tilde Q_i=\ell_s Q_i$, write
\be
g_{i \mu\nu}(s, x)= \tilde Q_i\, g_{\mu\nu}(t_i+\frac{s}{\tilde Q_i}, x)~,
\label{remetric2}
\ee
for $s\in [-\tilde Q_i t_i, 0]$ and take the limit as $i\rightarrow +\infty$.  The existence of the limit $g_\infty$ implies that the lengths as measured by
$g_\infty$ are finite at least for energy scales with $s\in (-\infty, 0]$ even though those measured by $g$ are infinite at $s=0$ and $t_i\rightarrow T$.

It has been shown that $g_\infty$ is a shrinking Ricci soliton \cite{Perelman, Naber, Enders}. As it has already  been mentioned, Ricci solitons solve the RG flow equation independently on whether they are thought as modelling the geometry of $M^n$ near a  type I singularity or not. But, thinking of $g_\infty$ as modelling the geometry of $M^n$ near a type I singularity,  its role in the context of perturbation theory for  sigma model with target space $(M^n, g)$ is not clear.  This is unless an interpretation is found of the limit $(M^n, g_i)$ within quantum field theory as $i\rightarrow \infty$. However, there are some clues. A sigma model with target space $(\tilde M^n, g_\infty)$ will ``ignore'' all the degrees of freedom that arise from far away from the singularity regions of $(M^n, g)$. As for the construction of $g_\infty$ both the original  metric $g$ and flow parameter $t$ are re-scaled, the degrees of freedom described by $g_\infty$ are those collected as the square of the coupling, $\ell_s^2/R_M^2$, goes to zero and the energy scale as represented by the flow parameter $t$ goes to infinity such that their ratio remains finite. It would be instructive, if a more intrinsic description can be found for these degrees of freedom but we have been unable to do so. For some of the cosmological applications below,  we shall make the assumption that the cosmological constant in the description of $g_\infty$ emerges as a cut off  for the degrees of freedom of the sigma model which are needed to describe the theory near the singularity.

\section{RG flows,  cosmology and some remarks}


To probe an application to cosmology from the results we have described so far, the main assumption that one has to make is that after the universe emerges from the quantum epoch, described by a theory of quantum gravity, and enters its geometric phase, described by general relativity type of theory, it satisfies a flow equation like (\ref{fl1}). Again note that (\ref{fl1}) does not exhibit a cosmological constant.

 This assumption alone allows for solutions that are  shrinking (or expanding) Ricci solitons. More generally, it allows for solutions  spacetimes  with a positive (or negative) cosmological constant. So far, we have described Ricci solitons in the context of manifolds of Euclidean signature but all the equations and arguments that do not involve limits, and so rely on the positive definiteness  property of norms, extend to manifolds with Lorentzian signatures.  In particular, the key argument that Ricci solitons can either been seen as solutions of (\ref{fl1}) or as solutions of (\ref{steadsol}) is valid for manifolds with Lorentzian signature. In particular, de-Sitter space, which is the typical solution that describes an expanding universe and with positive cosmological constant, is a solution of RG flow equation (\ref{fl1}), exhibiting a singularity at the IR.

 Let $g_{\mathrm dS}$ be the de-Sitter metric with cosmological constant $\Lambda$, $R(g_{\mathrm dS})_{\mu\nu}=\Lambda\, (g_{\mathrm dS})_{\mu\nu}$, $\Lambda>0$.  Then, the solution of the flow equation\footnote{In the context of cosmology, it should be stressed that $t$ should not be identified with a time coordinate. Instead, it is related to the energy scale of two dimensional sigma model.} is $g_{\mathrm dS}(t)=(1-\Lambda t) g_{\mathrm dS}$.  The Ricci tensor does not depend on the overall scale factor. So, we have
 \be
 R(g_{\mathrm dS}(t))_{\mu\nu}=R(g_{\mathrm dS})_{\mu\nu}= \Lambda\, (g_{\mathrm dS})_{\mu\nu}=\frac{\Lambda}{1-\Lambda t} (g_{\mathrm dS}(t))_{\mu\nu}~.
 \ee
 Thus the effective cosmological constant of the spacetime with metric $g_{\mathrm dS}(t)$ is $\Lambda(t)=\frac{\Lambda}{1-\Lambda t}$. Clearly, this is a de-Sitter metric for every $t$. Observe that $\Lambda(t)$ increases as $t<\Lambda^{-1}$ goes into IR. The flow parameter $t$ is related to the energy scale of the two-dimensional sigma model. But it is curious that   the cosmological constant is not a constant and depends non-trivially on $t$.

 So far, we have not introduced matter fields but this can be corrected by modifying the flow equations to include them. For example, one can write\footnote{The flow equation for the metric can be arranged in a different way using the Einstein tensor instead of the Ricci tensor. In the stated arrangement, the Ricci part has the diffusion and reaction terms as the Ricci flow and the matter contribution is treated as a reaction term, see \cite{Hamilton0}.}
 \bea
 \frac {d g(t)_{\mu\nu} }{dt} &=& - \Big(R(g(t))_{\mu\nu}- \kappa \tilde T(\phi(t), g(t))_{\mu\nu}\Big)~,
 \cr
 \frac {d \phi(t) }{dt} &=&- {\mathcal F}(\phi(t), \partial \phi(t), \partial^2 \phi(t), g(t))~,
 \label{fl2}
 \eea
 where $\tilde T$ is related to the energy momentum tensor $T$ of the field $\phi$ as $\tilde T_{\mu\nu}=T_{\mu\nu}-\frac{1}{2} g_{\mu\nu} T$, with $T=g^{\mu\nu} T_{\mu\nu}$,
 and ${\mathcal F}$ are the field equations of $\phi$. The field $\phi$ collectively denotes any other matter field of interest that one can consider.  Many of the arguments we have made for (\ref{fl1}) apply for the equations above but a more detailed exposition will be presented elsewhere.

 Stronger results for the cosmological constant can be obtained by making additional assumptions. For example, suppose that the spacetime as it emerges from the quantum epoch to its geometric phase has regions with positive and negative curvature.  In addition, it satisfies a flow equation like (\ref{fl1}).
 The regions with positive curvature will shrink and potentially lead to singularities.  Assuming that these singularities are described by a shrinking Ricci soliton\footnote{For type I singularities and Euclidean signature manifolds, this is a theorem. But the proof involves the use of limits and the proof may not be generalised to manifolds with Lorentzian signature metrics, see also \cite{Anderson}.} with metric $g_\infty$ and that $g_\infty$ describes the correct degrees of freedom to analyse the behaviour of the theory\footnote{This assumption has to be made independently on whether the spacetime has Euclidean or Lorentzian signature.}, then a positive cosmological constant emerges. As this can happen at different regions of spacetime with positive curvature, a model of the universe can arise that has different cosmological constants at different epochs.


Although there is no direct evidence, it is worth pointing out an analogy in the context of RG flows between Thurston's geometrisation conjecture and the homogeneity and isotropy of the universe.   Thurston's conjecture, now proven by Perelman, states that all 3-dimensional manifolds can be constructed by an appropriate gluing of eight homogeneous spaces.  The proof, as approached in the work of Hamilton and Perelman \cite{Hamilton0, Perelman2}, requires the continuation of the flow (\ref{fl1}) after a singularity is encountered at some finite $t=T$. This involves, at least for type I singularities, the identification of the geometry given by $g_\infty$. Then, surgery is performed to remove the part of the manifold that develops a singularity before the singularity is formed. Typically, in the place of the removed piece of the manifold with surgery   a disc is glued with carefully controlled curvature. Then, the flow is restarted and the procedure is repeated at every singularity until the geometry of the remaining manifolds is identified.  Thus, one can envisage the possibility to start from a universe with a generic geometry after the Big Bang and apply RG flows that evolves it to a homogeneous one. If such a calculation can be set up, it can potentially lead to a geometric proof of the homogeneity and isotropy of the universe.

\vskip1cm
 \noindent {\it {Acknowledgements}}

 I thank Kostas Skenderis and Andreas Stergiou  for helpful discussions.

 \bibliographystyle{unsrt}

\end{document}